\documentclass[aps,prd,nofootinbib]{revtex4}

\begin{document}
\title{The double role of Einstein's equations: as equations of motion
and as vanishing energy-momentum tensor\footnote{To Jerzy Pleba\'nski on the
occasion of his 75 the birthday.}}
\date{\today}

\author{Merced Montesinos}
\email{merced@fis.cinvestav.mx} \affiliation{Departamento de F\'{\i}sica,
Centro de Investigaci\'on y de Estudios Avanzados del I.P.N., Av. I.P.N. No.
2508, 07000 Ciudad de M\'exico, M\'exico.}

\begin{abstract}
Diffeomorphism covariant theories with dynamical background metric, like
gravity coupled to matter fields in the way expressed by Einstein-Hilbert's
action or relativistic strings described by Polyakov's action, have `on-shell'
vanishing energy-momentum tensor $t_{\mu\nu}$ because $t_{\mu\nu}$ is,
essentially, the Eulerian derivative associated with the dynamical background
metric and thus $t_{\mu\nu}$ vanishes `on-shell.' Therefore, the equations of
motion for the dynamical background metric play a double role: as equations of
motion themselves and as a reflection of the fact that $t_{\mu\nu}=0$.
Alternatively, the vanishing property of $t_{\mu\nu}$ can be seen as a
reflection of the so-called `problem of time' present in diffeomorphism
covariant theories in the sense that $t_{\mu\nu}$ are written as linear
combinations of first class constraints only.
\end{abstract}


\maketitle

\section{Introduction}
There are in literature several attempts to define the energy-momentum tensor
for gravity coupled to matter fields. One can say that there is, currently,
non agreement about how it must be defined or even if its definition makes
sense or not. The fact of having in the literature various definitions for it
is usually regarded as the reflection of the fact that this task makes no
sense:

\begin{center}
{\it `Anybody who looks for a magic formula for ``local gravitational
energy-momentum'' is looking for the right answer to the wrong question'}
\end{center}
is, for instance, a quotation found in page 467 of Ref. \cite{Misner}. In
spite of the strongness of that statement, the issue of the energy-momentum
tensor for gravity coupled to matter fields is investigated in this paper.
Intuitively, it is expected that all dynamical fields of which an action
depends on contribute to the full energy-momentum tensor for the full system
of dynamical fields. So, from this perspective it is natural to expect a
contribution of the gravitational field to the full energy-momentum tensor.
The viewpoint adopted here about the definition of the energy-momentum tensor
is, from the mathematical point of view, very simple. However, it will be more
important for us to explore the conceptual aspects involved in such a
definition. Specifically, the definition of the energy-momentum tensor
$t_{\mu\nu}$ is taken as that coming from the variation of the action under
consideration with respect to the dynamical metric\footnote{It might be
possible to consider the first order formalism if fermions want to be
included. The ideas developed here can without any problem be applied to that
case.} $g_{\mu\nu}$ the action depends on
\begin{eqnarray}
t_{\mu\nu} := \frac{2}{\sqrt{-g}} \frac{\delta S[g_{\mu\nu}, \phi ]}{\delta
g^{\mu\nu}} \, . \label{def}
\end{eqnarray}
From the definition of Eq. (\ref{def}) it is evident that, for a dynamical
background metric, the energy-momentum tensor $t_{\mu\nu}$ vanishes `on-shell'
\begin{eqnarray}
t_{\mu\nu} = 0 \, .
\end{eqnarray}
In particular, in the case of gravity coupled to matter fields $\phi$,
$t_{\mu\nu} = T_{\mu\nu}- \frac{c^4}{8\pi G} G_{\mu\nu}$, which vanishes
because of Einstein's equations. To `avoid this difficulty' (i.e., the
`on-shell' vanishing property of $t_{\mu\nu}$) people usually say that this
way of defining the energy-momentum tensor just gives the `right' form for the
energy-momentum tensor of the matter fields $\phi$ only, $T_{\mu\nu}$, by
simply identifying in the expression for $t_{\mu\nu}$ the contribution of the
matter fields $\phi$. We disagree with that point of view because it puts the
matter fields $\phi$ and the geometry $g_{\mu\nu}$ on non the same footing
(i.e., the status of the matter fields $\phi$ is (from that perspective)
different to the status of the geometry $g_{\mu\nu}$ even though both fields
are dynamical ones). Here, on the other hand, we argue that there is nothing
wrong either with the definition of the energy-momentum tensor $t_{\mu\nu}$
given in Eq. (\ref{def}) nor with the fact that it vanishes `on shell.' This
way of interpreting things has several conceptual consequences:

\begin{enumerate}
\item
all fields the action depends on are on the same conceptual footing and thus
it is not a surprise that they all contribute (as dynamical fields) to the
full energy-momentum tensor $t_{\mu\nu}$.
\item
a natural definition for the energy-momentum tensor of the gravitational field
arises.
\item
the vanishing property of $t_{\mu\nu}$ is not interpreted as a `problem' which
must be corrected somehow but rather as a reflection of the double role that
the equations of motion associated with the dynamical background play or,
alternatively, as a reflection of diffeomorphism covariance.
\end{enumerate}

The content of the paper is organized as follows. In Sect. 2, the issue of the
energy-momentum tensor for Polyakov's action is studied where point (3) is
explicitly displayed. Next, in Sect. 3, the system of gravity coupled to
matter fields is analyzed. Finally, our conclusions are collected in Sect. 4.

\section{Polyakov's action}
Before considering general relativity coupled to matter fields, and as a
warming up, let us study the dynamics of relativistic bosonic strings
propagating in an arbitrary $D$-dimensional fixed (i.e., non dynamical)
background spacetime with metric $g=g_{\mu\nu} (X) d X^{\mu} d X^{\nu}$;
$\mu,\nu=0,1,...,D-1$. This system can be described, for instance, with the
action\cite{Brink,Deser,Polyakov}
\begin{eqnarray}
S[\gamma^{ab}, X^{\mu}] = \alpha \int_{\cal M} d^2 \xi \sqrt{-\gamma}
\gamma^{ab} \partial_a X^{\mu} \partial_b X^{\nu} g_{\mu\nu} (X) \, .
\label{Polyakov}
\end{eqnarray}

The variation of the action (\ref{Polyakov}) with respect to the background
coordinates $X^{\mu}$ and the inverse metric $\gamma^{ab}$ yields the
equations of motion
\begin{eqnarray}
\nabla^a \nabla_a X^{\mu} + \Gamma^{\mu}\,_{\alpha\beta} \,\, \gamma^{bc}
\partial_b X^{\alpha} \partial_c X^{\beta} & = & 0 \, , \label{EqM1}\\
\frac{\alpha}{2} \gamma_{ab} \gamma^{cd} h_{cd}- \alpha h_{ab} =0 \, ,
\label{EqM2}
\end{eqnarray}
with
\begin{eqnarray}
h_{ab} & = & \partial_a X^{\mu} \partial_b X^{\nu} g_{\mu\nu} (X) \, ,
\end{eqnarray}
the induced metric on the world sheet ${\cal M}$.

{\it Energy-momentum tensor}. The energy-momentum tensor $t_{ab}$ for the full
system of fields is defined by the variational variation of the action
(\ref{Polyakov}) with respect to the inverse metric $\gamma^{ab}$
\begin{eqnarray}
t_{ab} & := &  \frac{2}{\sqrt{-\gamma}}\frac{\delta S[\gamma^{ab},
X^{\mu}]}{\delta \gamma^{ab}}
\nonumber\\
 & = & \frac{\alpha}{2} \gamma_{ab} \gamma^{cd} h_{cd}- \alpha h_{ab} \, .
 \label{tensor}
\end{eqnarray}
Thus, $t_{ab}$ is defined with respect to any `local' observer on the world
sheet ${\cal M}$. Note that the trace of $t_{ab}$ vanishes `off-shell,'
$t_{ab} \gamma^{ab}=0$. Moreover, by using Eq. (\ref{EqM2})
\begin{eqnarray}
t_{ab} = 0 \, ,
\end{eqnarray}
i.e., the energy-momentum tensor vanishes `on-shell' or, what is the same,
$t_{ab}=0$ is just a reflection of having a dynamical background metric on the
world sheet ${\cal M}$. In this sense, Eq. (\ref{EqM2}) plays a {\it double
role}. The first role of Eq. (\ref{EqM2}) is that it is the equation of motion
associated with the dynamical field $\gamma^{ab}$. The second role of Eq.
(\ref{EqM2}) is that it is the reflection of the fact that the energy-momentum
tensor of the full system vanishes, $t_{ab}=0$, which in fact implies that
$\gamma_{ab}$ is proportional to the induced metric $h_{ab}$, $\gamma_{ab}=
e^{\Omega} h_{ab}$. Alternatively, the fact that the energy-momentum tensor
$t_{ab}$ vanishes from the viewpoint of an observer sitting on the world sheet
establishes a {\it balance} between the intrinsic metric $\gamma_{ab}$ and the
induced metric $h_{ab}$ in the precise form given by $\gamma_{ab}=e^{\Omega}
h_{ab}$.

{\it Diffeomorphism covariance (active diffeomorphism invariance)}.
Diffeomorphism covariance or general covariance from the {\it active} point of
view means the following. Let $(X^{\mu} (\xi), \gamma_{ab} (\xi))$ be {\it any
solution} of the equations of motion (\ref{EqM1}) and (\ref{EqM2}) with
respect to the coordinate system associated with the local coordinates $\xi$
and let $f:{\cal M}\rightarrow {\cal M}$ be any diffeomorphism of the world
sheet ${\cal M}$ onto itself then, the new configuration
\begin{eqnarray}
{X'}^{\mu} (\xi) & = &  X^{\mu} (f(\xi)) \, , \nonumber\\
{\gamma'}_{ab} (\xi)  & = & \frac{\partial f^c}{\partial \xi^a} \frac{\partial
f^d}{\partial \xi^b} \gamma_{cd} (f(\xi)) \, , \label{PolDiff}
\end{eqnarray}
is also a (mathematically different) solution to the equations of motion
(\ref{EqM1}) and (\ref{EqM2}) with respect to the {\it same} observer. Even
though, $(X^{\mu} (\xi), \gamma_{ab} (\xi))$ and $({X'}^{\mu} (\xi),
{\gamma'}_{ab} (\xi))$ are mathematically distinct configurations they
represent the {\it same} physical solution with respect to the local observer,
i.e., any diffeomorphism of the world sheet ${\cal M}$ induces a gauge
transformation on the fields living on ${\cal M}$ which is given in Eq.
(\ref{PolDiff}). As it is well-known, gauge symmetries are, in the canonical
formalism, associated with first class constraints.\cite{Dirac}

{\it $1+1$ canonical viewpoint of the energy-momentum tensor}. The vanishing
of $t_{ab}$ is just a reflection of the gauge symmetry (\ref{PolDiff}). To see
this, ${\cal M}=R\times \Sigma$ and the metric $\gamma_{ab}$ is put in the ADM
form
\begin{eqnarray}
\left ( \gamma_{ab} \right ) =  \left (
\begin{array}{ll}
  - N^2 + \lambda^2  \chi & \chi  \lambda \\
\,\,\,\,\,\chi \lambda & \chi
\end{array}
\right )\, , \nonumber\\
\left ( \gamma^{ab} \right ) =  \left (
\begin{array}{ll}
-\frac{1}{N^2} & \frac{\lambda}{N^2}\\
\,\,\,\,\,\frac{\lambda}{N^2} & \frac{1}{\chi} - \frac{\lambda^2}{N^2}
\end{array}
\right )\, , \label{ADM}
\end{eqnarray}
and so $\sqrt{-\gamma} := \sqrt{-\det{\gamma_{ab}}} = \epsilon N \sqrt{\chi}$
with $\epsilon=+1$ if $N>0$ and $\epsilon=-1$ if $N<0$. Due to the fact
$\frac{\partial}{\partial \tau}$ is time-like and $\frac{\partial}{\partial
\sigma}$ is space-like then $- N^2 + \lambda^2 \chi < 0$ and $\chi>0$. Taking
into account Eq. (\ref{ADM}), the action of Eq. (\ref{Polyakov}) acquires the
form
\begin{eqnarray}
S[X^{\mu}, P_{\mu}, M , \lambda ] = \int_R d\tau \int_{\Sigma} d \sigma \left
[ {\dot X}^{\mu} P_{\mu} - \left ( M H + \lambda D \right ) \right ]\, ,
\label{Haction}
\end{eqnarray}
with
\begin{eqnarray}
H  & := & P_{\mu} P_{\nu} g^{\mu\nu} + 4 \alpha^2 {X'}^{\mu} {X'}^{\nu}
g_{\mu\nu} \, , \nonumber\\
D  & := & {X'}^{\mu} P_{\mu} \, . \label{constII}
\end{eqnarray}
The dependence of the phase space variables and Lagrange multipliers in terms
of the Lagrangian variables is
\begin{eqnarray}
P_{\mu} & := & -\frac{2 \alpha \epsilon \sqrt{\chi}}{N} {\dot X}^{\nu}
g_{\mu\nu} + \frac{2\alpha \epsilon \lambda \sqrt{\chi}}{N} {X'}^{\nu}
g_{\mu\nu} \, , \nonumber\\
M & := & -\frac{N}{4 \alpha\epsilon\sqrt{\chi}} \, , \label{DEF}
\end{eqnarray}
where ${X'}^{\mu} = \frac{\partial X^{\mu}}{\partial \sigma}$.

Hamilton's principle applied to the action (\ref{Haction}) yields the
dynamical equations
\begin{eqnarray}
{\dot X}^{\mu} & = & 2 M P_{\nu} g^{\mu\nu} + \lambda {X'}^{\mu} \, ,
\label{Dyn1}\\
{\dot P}_{\mu} & = & M \, Y^{\theta\phi} \frac{\partial
g_{\theta\phi}}{\partial X^{\mu} } + \left ( 8 \alpha^2 M {X'}^{\nu}
g_{\mu\nu} + \lambda P_{\mu} \right )'\, , \label{EMOTION}
\end{eqnarray}
and
\begin{eqnarray}
H \approx 0 \, , \quad D \approx 0\, , \label{Csurface}
\end{eqnarray}
which are the Hamiltonian and diffeomorphism first class constraints;
respectively. Here, $Y^{\theta\phi} = P_{\mu} P_{\nu} g^{\theta\mu}
g^{\phi\nu} - 4 \alpha^2 {X'}^{\theta} {X'}^{\phi}$.

Next, the induced metric $h_{ab}$ in terms of the phase space variables and
Lagrange multipliers is written down
\begin{eqnarray}
h_{\tau\tau} & = & {\dot X}^{\mu} {\dot X}^{\nu} g_{\mu\nu} \nonumber\\
& = & 4 M^2 P_{\mu} P_{\nu} g^{\mu\nu} + 4 M \lambda P_{\mu} {X'}^{\mu} +
\lambda^2 {X'}^{\mu} {X'}^{\nu} g_{\mu\nu}\, , \nonumber\\
h_{\tau\sigma} & = & {\dot X}^{\mu} {X'}^{\nu} g_{\mu\nu} \nonumber\\
& = & 2 M P_{\mu} {X'}^{\mu} +
\lambda {X'}^{\mu} {X'}^{\nu} g_{\mu\nu} \, , \nonumber\\
h_{\sigma\sigma} & = &  {X'}^{\mu} {X'}^{\nu} g_{\mu\nu} \, , \label{Aux}
\end{eqnarray}
where the dynamical equation (\ref{Dyn1}) was used.

Therefore, by using Eqs. (\ref{tensor}), (\ref{ADM}), (\ref{Aux}), and the
definition of the constraints (\ref{constII}), the components of the
energy-momentum tensor (\ref{tensor}) in terms of the phase space variables
and the Lagrange multipliers acquire the form\cite{MerVer}
\begin{eqnarray}
t_{\tau\tau} & = & -2\alpha M^2 \left ( 1 + \frac{\lambda^2}{16 \alpha^2 M^2 }
\right ) H - (4 \alpha M \lambda )\, D\, , \nonumber\\
t_{\tau\sigma} & = & -\frac{\lambda}{8\alpha} H
- (2 \alpha M) \, D \, , \nonumber\\
t_{\sigma\sigma} & = & -\frac{1}{8\alpha} H \, . \label{key}
\end{eqnarray}
Thus, Eq. (\ref{key}) clearly expresses the conceptual reason of the vanishing
property of $t_{ab}$, i.e., $t_{ab}$ vanish because they are (modulo the
dynamical Eq. (\ref{Dyn1})) linear combinations of the first class constraints
(\ref{Csurface}) for the system, which are the $1+1$ canonical version of the
gauge symmetry (\ref{PolDiff}). In conclusion, a vanishing energy-momentum
tensor $t_{\mu\nu}=0$ is a reflection of the fact that the hamiltonian of the
theory is just a linear combination of first class constraints. One could say
that $t_{\mu\nu}=0$ is equivalent to the definition of the `constraint
surface', however, this is not so because Eq. (\ref{key}) was written by using
the dynamical equation (\ref{Dyn1}).

\section{Gravity coupled to matter fields}
Now, let us study the Einstein-Hilbert action coupled to matter fields
\begin{eqnarray}
S [g_{\mu\nu} , \phi] & = & \frac{c^3}{16\pi G}\int_{\cal M} \sqrt{-g}\,\, R
\,\, d^4 x + \int_{\cal M} \sqrt{-g} \,\, L_{matter \hspace*{2pt} fields}
(\phi) \,\, d^4 x \, , \label{fullA}
\end{eqnarray}
where $R$ is the scalar curvature and $L_{matter \hspace*{2pt} fields}(\phi)$
denotes the contribution of the matter fields dynamically coupled to gravity
and denoted generically by $\phi$.

Hamilton's principle yields the equations of motion for the system
\begin{eqnarray}
\frac{\delta S}{\delta \phi} & = & 0 \, , \\
G_{\mu\nu} & = & \frac{8 \pi G}{c^4} T_{\mu\nu} \, , \label{E1}
\end{eqnarray}
where $G_{\mu\nu}= R_{\mu\nu} - \frac12 R g_{\mu\nu}$ is Einstein's tensor and
$T_{\mu\nu}$ is the contribution of the matter fields $\phi$ to the
energy-momentum tensor.\footnote{Strictly speaking, the field $g_{\mu\nu}$
also contributes to $T_{\mu\nu}$.}

{\it Energy-momentum tensor}. Again, the energy-momentum tensor $t_{\mu\nu}$
is obtained by the variation of the full action (\ref{fullA}) with respect to
the inverse metric $g^{\mu\nu}$. Therefore, the energy-momentum tensor for the
full system of fields is
\begin{eqnarray}
t_{\mu\nu} = T_{\mu\nu} - \frac{c^4}{8\pi G} G_{\mu\nu} \, . \label{emgrav}
\end{eqnarray}
Some remarks follow:

1) from this perspective, $T_{\mu\nu}$ is the contribution of the matter
fields $\phi$ while $- \frac{c^4}{8\pi G} G_{\mu\nu}$ is the contribution of
the gravitational field $g_{\mu\nu}$ to $t_{\mu\nu}$. Therefore, the matter
fields $\phi$ and the gravitational field $g_{\mu\nu}$ are put on the {\it
same} ontological status in the sense that both of them contribute (as
dynamical fields) to the full energy-momentum tensor $t_{\mu\nu}$. In
addition, and in contrast to the dynamical system described by Polyakov's
action, note that the full energy-momentum tensor of gravity and matter fields
given in Eq. (\ref{emgrav}) is composed of two additive parts each one being
associated to each field, i.e., there is a splitting of the contributions of
the fields ($g_{\mu\nu}$ and $\phi$) to $t_{\mu\nu}$.\footnote{See footnote
b.}

2) for observers which detect a gravitational field the energy-momentum tensor
identically vanishes, $t_{\mu\nu}=0$, because of Einstein's equations
(\ref{E1}). This means that for this type of observers, there is a {\it
balance} between the `content' of energy and momentum densities and stress
associated with the matter fields $\phi$ (which is characterized in
$T_{\mu\nu}$) and the `content' of energy and momentum densities and stress
associated with the gravitational field (which is characterized in $-
\frac{c^4}{8\pi G} G_{\mu\nu}$)
\begin{eqnarray}
\longrightarrow \longrightarrow \longrightarrow \longrightarrow  \nonumber\\
\longleftarrow \longleftarrow \longleftarrow \longleftarrow
\end{eqnarray}
in a precise form such that both fluxes cancel, and thus leading to a
vanishing `flux', i.e., $t_{\mu\nu}=0$. Once again, the vanishing property of
$t_{\mu\nu}$ for the system of gravity coupled to matter fields is just a
reflection of the fact that the background metric is dynamical. More
precisely, $t_{\mu\nu}=0$ tells us that the `reaction' of the dynamical
background metric is such that it just cancels the effect of `flux' associated
with the matter fields. It is impossible (and makes no sense) to have a
locally non-vanishing `flux' in this situation. If this were the case, there
would be no explanation for the origin of that non-vanishing `flux'. Moreover,
that hypothetic non-vanishing `flux' would define privileged observers
associated with it (the ether would come back!). It is important to emphasize
that, in the case of having a dynamical background metric, the vanishing
property of $t_{\mu\nu}$ is {\it not} interpreted here as a `problem' that
must be corrected somehow but exactly the other way around. In our opinion,
there is nothing wrong with that property because it just reflects the double
role that the equations of motion associated with the dynamical background
play.

3) Connection with special relativity. In the conceptual framework of the
special theory of the relativity of motion the background metric is {\it
fixed} (i.e., non-dynamical\footnote{This is also true for any field theory
defined on a curved fixed background, however, this is not relevant for the
present purposes.}), the only dynamical entities are the matter fields $\phi$
and thus any Lorentz observer can associate a non-vanishing `content' of
energy and momentum densities and stress
\begin{eqnarray}
\longrightarrow \longrightarrow \longrightarrow \longrightarrow \nonumber\\
\longrightarrow \longrightarrow \longrightarrow \longrightarrow
\end{eqnarray}
associated to them (the fields) and represented in $T_{\mu\nu}$. How does this
non-vanishing $t_{\mu\nu}=T_{\mu\nu}$ in the special theory of the relativity
of motion come out from general relativity where $t_{\mu\nu}=0$? If one goes
from the general to the special theory of the relativity of motion by using
`locally' freely-falling observers one finds that there is a contradiction
between the fact explained above and the point 2) which states that
$t_{\mu\nu}=0$ must hold for {\it any} observer and, in particular, for
`locally' freely-falling observers which find `no evidence of gravity'
(Einstein's equivalence principle). However, the contradiction disappears by
noting that in the point 2) the background field metric is dynamical while in
special relativity is not, i.e., the contradiction arises from the comparison
of two conceptually different scenarios.

More precisely, the fact of `locally' having the special theory of the
relativity of motion and therefore a non-vanishing energy-momentum tensor
(whose contribution comes only from the matter fields) is just a reflection of
{\it neglecting} (by means of Einstein's equivalence principle) the
contribution of the dynamical background $g_{\mu\nu}$ to the full tensor,
i.e., `locally' freely-falling observers can {\it not} use Einstein's
equations simply because for them the background metric is {\it non dynamical}
but it is {\it fixed} to be the Minkowski metric. What these observers do is
simply to neglect the second term in the right-hand side of Eq. (\ref{emgrav})
under the pretext of Einstein's equivalence principle.  Note, however, that
from the mathematical point of view it is {\it not} possible to do that
because it is impossible to choose local coordinates (and thus a particular
reference frame attached to it) such that with respect to these coordinates
(with respect to this reference frame) the Riemann tensor
$R_{\alpha\beta\gamma\delta}$ vanishes in a certain point (in whose
neighborhood one could define the concept of an `inertial reference frame' in
the sense of the conceptual framework of the special theory of the relativity
of motion). This mathematical impossibility is other way of saying that
particular reference frames where the gravitational field (represented by the
Riemann tensor) completely vanishes do {\it not} exist. This fact implies that
it is {\it impossible} to cancel the effects of the gravitational field {\it
even} for freely-falling observers because of the presence of tidal forces.
Therefore, it is conceptually not possible to neglect gravity effects and thus
all observers must conclude that the background metric is always dynamical and
that its effects can not be neglected. Thus, conceptually, $t_{\mu\nu}=0$
always. If, by hand (Einstein's equivalence principle) the dynamics of the
background metric is neglected then this fact leads to the arising of
non-vanishing energy-momentum tensor associated with matter fields only.

\section{Concluding Remarks}
The lesson from Polyakov's action and from gravity coupled to matter fields
leaves no room for speculations. It is completely clear the relationship
between diffeomorphism covariance and a vanishing energy-momentum tensor
$t_{\mu\nu}$ in both theories. Alternatively, one could say that the vanishing
property of $t_{\mu\nu}$ is another manifestation  of the so-called `the
problem of time' which, of course, is not a problem but a property of
generally covariant theories. Moreover, the interplay between $t_{\mu\nu}$ and
the Euler-Lagrange derivative associated with the dynamical background metric
in the way expressed in Eq. (\ref{emgrav}) leaves no room for attempts of
modifying the expression for the energy-momentum tensor adding, for instance,
divergences because if this were done, say, that a hypothetic `right'
energy-momentum tensor ${\cal T}_{\mu\nu}$ were built
\begin{eqnarray}
{\cal T}_{\mu\nu} & = & t_{\mu\nu} + \nabla_{\gamma} \chi_{\mu\nu}\,^{\gamma}
\nonumber\\
& = & T_{\mu\nu} - \frac{c^4}{8\pi G} G_{\mu\nu} + \nabla_{\gamma}
\chi_{\mu\nu}\,^{\gamma} \, , \label{impten}
\end{eqnarray}
by this procedure then, ${\cal T}_{\mu\nu}=0$ would imply
\begin{eqnarray}
G_{\mu\nu} & = & \frac{8\pi G}{c^4} \left ( T_{\mu\nu} + \nabla_{\gamma}
\chi_{\mu\nu}\,^{\gamma} \right )\, ,
\end{eqnarray}
thus modifying the original Einstein's equations (\ref{E1}) we start with
which is a contradiction, i.e., any attempt to `improve' the energy-momentum
by adding divergence terms, $t_{\mu\nu} \rightarrow {\cal T}_{\mu\nu}$, would
modify the field equations associated with the background metric and there is
currently no experimental reason to do that. So, `improvements' for the
energy-momentum tensor $t_{\mu\nu}$ of the kind introduced by Belinfante are
not allowed in diffeomorphism covariant theories.

As a final comment let us consider the theory of a massless scalar field
defined on a flat background expressed in a generally covariant
form\cite{Sorkin}
\begin{eqnarray}
S[g_{\mu\nu}, \phi, \lambda^{\mu\nu\gamma\delta} ] & = & -\frac12 \int_{\cal
M} \sqrt{-g} g^{\mu\nu} \partial_{\mu} \phi \partial_{\nu} \phi + \frac14
\int_{\cal M} \sqrt{-g} \lambda^{\mu\nu\gamma\delta} R_{\mu\nu\gamma\delta} \,
. \label{SFGC}
\end{eqnarray}
Hamilton's principle yields the equations of motion
\begin{eqnarray}
\nabla^{\gamma} \nabla^{\delta} \lambda_{\mu\gamma\nu\delta} & = &
T_{\mu\nu}\, , \label{A}\\
g^{\mu\nu} \nabla_{\mu} \nabla_{\nu} \phi & = & 0 \, , \label{B}\\
R_{\mu\nu\gamma\delta} & = & 0 \, . \label{C}
\end{eqnarray}
Note that the first equation plays the role of Einstein's equations, i.e., it
is the equation associated with the dynamical background metric.\footnote{The
background metric $g_{\mu\nu}$ is dynamical in the sense that the action
(\ref{SFGC}) depends functionally on it.} Again, the energy momentum tensor
for the system is
\begin{eqnarray}
t_{\mu\nu} = T_{\mu\nu} - \nabla^{\gamma} \nabla^{\delta}
\lambda_{\mu\gamma\nu\delta} \, ,
\end{eqnarray}
and it vanishes because of Eq. (\ref{A}). Therefore, one finds the same
phenomenon found in gravity coupled to matter fields in the sense that if
`locally' freely-falling observers {\it neglected} the reaction of the
background (i.e., neglecting $\nabla^{\gamma} \nabla^{\delta}
\lambda_{\mu\gamma\nu\delta}$) they would observe just a non-vanishing
$t_{\mu\nu}= T_{\mu\nu}$, as expected. However, note that {\it conceptually}
(i.e., from the mathematical point of view) it is impossible to neglect
$\nabla^{\gamma} \nabla^{\delta} \lambda_{\mu\gamma\nu\delta}$ because it is
not possible to find a coordinate system and a point in which this term
vanishes. Note also that the theory defined by the action of Eq. (\ref{SFGC})
is completely different to the theory defined by
\begin{eqnarray}
S[\phi] = -\frac12 \int_{\cal M} \sqrt{-g} g^{\mu\nu} \partial_{\mu} \phi
\partial_{\nu} \phi \, . \label{NNN}
\end{eqnarray}
(assuming that the background metric $g_{\mu\nu}$ is flat) in the following
sense: in the field theory defined in Eq. (\ref{SFGC}), $t_{\mu\nu}=0$ because
of the dynamical equation for $g_{\mu\nu}$ while in the field theory defined
in Eq. (\ref{NNN}) the background metric $g_{\mu\nu}$ is non-dynamical and
thus the theory has a non-vanishing energy-momentum tensor
$t_{\mu\nu}=T_{\mu\nu}$. Of course, what defines a theory is its equations of
motion, so one could say whether or no both theories are the same by simple
looking at their equations of motion. However, from the present analysis, they
have different full energy-momentum tensors and this fact indicates that each
of these theories describe physically distinct scenarios. Moreover, the theory
defined by Eq. (\ref{NNN}) has, from the canonical point of view, a
non-vanishing Hamiltonian while the Hamiltonian for the theory defined by Eq.
(\ref{SFGC}) must involve first class constraints because of diffeomorphism
covariance. In summary:
\begin{eqnarray}
\begin{tabular}{|c|c|c|}
\hline & \mbox{{\small Theory of Eq. (\ref{SFGC})}}  & \mbox{{\small Theory
of Eq. (\ref{NNN})}} \\
\hline \mbox{{\small general covariance (passive diff. inv.) }} & Yes & Yes \\
\hline \mbox{{\small dynamical background metric}} & Yes & Not \\ \hline
\mbox{{\small diff. covariance (active diff. inv.)}} & Yes & Not \\ \hline
\mbox{{\small vanishing $t_{\mu\nu}$}} & Yes & Not \nonumber \\ \hline
\end{tabular}
\end{eqnarray}

\section*{Acknowledgments}
Warm thanks to U. Nucamendi and G.F. Torres del Castillo for useful comments
on the subject of this paper. The author also thanks support provided by the
{\it Sistema Nacional de Investigadores} (SNI) of the {\it Consejo Nacional de
Ciencia y Tecnolog\'{\i}a} (CONACyT) of Mexico.



\begin{thebibliography}{99}
\bibitem{Misner}
C. W. Misner, K. S. Thorne and J. A. Wheeler, {\it Gravitation} (W. H. Freeman
and Company, New York, 1973).
\bibitem{Brink}
L. Brink, P. Di Vechia and P. Howe, Phys. Lett. {\bf B65}, 471 (1976).
\bibitem{Deser}
S. Deser and B. Zumino, Phys. Lett. {\bf B65}, 369 (1976).
\bibitem{Polyakov}
A. M. Polyakov, Phys. Lett. {\bf B103}, 207 (1981); Phys. Lett. {\bf B103},
211 (1981).
\bibitem{Dirac}
P. A. M. Dirac, {\it Lectures on Quantum Mechanics} (Dover Publications, Inc.
Mineola, New York, 2001).
\bibitem{MerVer}
M. Montesinos and J. D. Vergara, Rev. Mex. F\'{\i}s. {\bf 40} {\bf S1}, 53
(2003); hep-th/0105026.
\bibitem{Sorkin}
R. D. Sorkin, Mod. Phys. Lett. A {\bf 17}, 695 (2002).
\end{thebibliography}
\end{document}